\documentstyle[12pt]{article}

\begin{document}

\author{Jos\'e I. Usera  \\ 
Universidad Complutense de Madrid}
\title{An Approach to Measurement by Quantum-Stochastic-Parameter Averaged Bohmian
Mechanics}
\date{1, 15, 2000}
\maketitle

\begin{abstract}
A coarse-grained quantum operator technique is used along with the formalism
of Bohmian mechanics endowed with stochastic character at the quantum level
in order to address some central issues in the quantum theory of
measurement. A surprisingly simple picture of decoherence and EPR\
correlations emerges from its use.
\end{abstract}

{\it ``It is the customary fate of new truths to begin as heresies and to
end as superstitions''}

Thomas\ Henry Huxley: {\it The Coming of Age of the Origin of Species.}

\section{Introduction}

The present work is devoted to the consistent removal from the quantum
formalism of the postulate where it gets closest sheer contradiction; the
so-called projection postulate. In its version for non-degenerate pure
states, it prescribes that immediately after a measurement with outcome $q$,
the quantum state must change as follows:

\begin{equation}
\label{collapse}\left| \psi (t_2)\right\rangle =\left\langle \psi
(t_1)\left| P_q\right| \psi (t_1)\right\rangle ^{-1/2}P_q\left| \psi
(t_1)\right\rangle 
\end{equation}
where $P_q=\left| q\right\rangle \left\langle q\right| $. Now, it is
impossible to obtain (\ref{collapse}) from the general quantum law of
evolution\ \qquad \ 
\begin{equation}
\label{Schrodinger-evolution}\left| \psi (t_2)\right\rangle
=U(t_2,t_1)\left| \psi (t_1)\right\rangle 
\end{equation}
for whatever linear and unitary $U(t_2,t_1)$, as (\ref{collapse}) is
nonlinear in its vector argument $\left| \psi (t_1)\right\rangle $. One
would try to escape contradiction by considering the denominator in (\ref{collapse})
 as a mere {\it convention} introduced in order to keep handy the
statistical interpretation of quantum mechanics for individual systems even
through the ocurrence of such accidents as ``measurements''. The problem is
the natural relaxation of (\ref{collapse})

\begin{equation}
\label{projection}\left| \psi (t_2)\right\rangle =P_q\left| \psi
(t_1)\right\rangle 
\end{equation}
is {\it also} an impossibility in combination with (\ref{Schrodinger-evolution})
 because of its non-unitary character. Put bluntly:
Copenhagen quantum mechanics (CQM) makes room for a precarious consistency
thanks to an artificial decree for two contradictory assumptions to hold at
different times and in different (but otherwise unspecified) contexts, thus
getting the nearest to logical ``collapse'' a theory has ever been. But we
are speaking of mathematical statements, so the point is mathematical
consistency, not philosophical fogginess.

It is perhaps ironical that all attempts to complete quantum mechanics\ have
been haunted for decades by the issuing of different proofs, all of them
consisting on some {\it reductio ad absurdum} argument, when this dazzling
absurdity already is among Copenhagen's quantum postulates. Such
counter-arguments have always proved extremely fragile, as showed by J. S.
Bell concerning Von Neumann's infamous impossibility theorem. The reason, no
doubt, is all of them are chained to one and the same burden of an
excessively close identification between elements of reality or {\it %
pre-existing properties of a system} and {\it eigenvalues of certain
operator.} The latter are only present in the linear level of the theory
(already thinking in Bohmian terms,) which is precisely the heaviest burden
one is relieved from when trying to extend quantum mechanics with additional
variables, and the former are only present in our unredeemably deterministic
minds. It is {\it only if} we (willingly) restrain ourselves from going
beyond the quantum-operator level that we find our hands tied when trying to
make a consistent and exhaustive (including every conceivable experiment)
model of the world. It seems pertinent to point out that outcomes of
experiments {\it are} always particle collisions in screens or counters, and
not operator eigenvalues. The fact that there is one possible detection for
each operator eigenvalue we can think of, must be looked upon as a theorem
of the additional variable\footnote{%
I intently avoid the malignant use of the word ``hidden'' for variables as
obvious as the coordinates of a pixel on a screen.} theory. And in this
additional-variable theory there is no need for elements of reality to
pre-exist before the experiment is performed because the theory may well be 
{\it contextual}. This forces, not only to feed $\left| \psi \right\rangle $
into the Kochen-Specker theorem \cite{Bell1}, \cite{Kochen-Specker}, \cite{Redhead}
 as part of the overall set of dynamical variables giving rise to
the particular outcomes, but also the potential energy, which is related to
the environment's dynamical state in a highly non-trivial way. In
particular, discontinuity in the potential energy both in time and space and
stochasticity may be involved in an essential way, and we know continuity
and determinism are both essential ingredients of the Kochen-Specker
argument (we will get back to this point in the concluding remaks).

In this work, I am facing the fact that neither (\ref{collapse}) nor (\ref{projection})
 are compatible with (\ref{Schrodinger-evolution}). In
particular, I find that, when Bohmian mechanics is added to the quantum
formalism, it{\it \ is }possible to contemplate measurements in a consistent
way substituting formula (\ref{collapse}) by

\begin{equation}
\label{multi-projection}\left| \psi (t_2)\right\rangle =\sum_{\forall q\in
\sigma (Q)}e^{-i\alpha _q}P_q\left| \psi (t_1)\right\rangle 
\end{equation}
where $\alpha _q$ are parameters both (i): stochastic in character, and
(ii): formally included within the quantum formalism so that (\ref{multi-projection})
 comes from a certain resolute approximation in the
Schr\"odinger equation. Of course, it will be necessary to complement the
resulting image with additional variables in order to supply for the
pointers which (\ref{multi-projection}) obviously lacks or, in other words,
to supply for the definiteness of the observed world. The essentials of this
approach were already developed by Bohm\cite{Bohm},\cite{Wheeler-Zurek}, and
there are other precedents in the literature as concerns the assumption of
the existence of a source of stochasticity at the quantum-dynamical level%
\cite{Namiki-Pascazio},\cite{Danieri-Loinger-Prosperi}. What, I gather, is
new in this approach\cite{Usera} is the realisation that formula (\ref{multi-projection})
 for different elections of the operator $Q$, fits at
least three paradigmatic situations in experiments performed with single
particles: (1$^{{\rm {st}}}$): collimation, (2$^{{\rm {nd}}}$): deflection
and (3$^{{\rm {rd}}}$): localisation.

To summarise, I will assume the following:

1$^{{\rm {st}}}$: I do not set out to develop an encompassing description of
the workings of the macroscopic world. I will adopt a much less ambitious
program instead, concerning such things as ``experiments''. For all their
narrowness, such situations do have a valuable advantage: they are very much
under human control. Thus, it is not a description of conscience or
particles in a thermal bath etc., I am concerned with. It is Stern-Gerlach,
EPR, double-slit, etc., {\it experiments}. There is, of course, an
undeniable interest in crossing over from these to more general realms. My
hope, however, is that once an elementary but consistent account of the
relevant topics for simple, concrete, {\it repeatable} instances is
presented, it will not be very difficult to believe that similar things must
happen in a more general context (after all, features such as coherence are
an exception, not a general rule).

The most important characteristic of these simplest of situations we call
Stern-Gerlach, EPR, etc. is {\it not} that they are made up of few dynamical
variables (which is actually not the case because macroscopic field sources
stem from the presence of swarms of particles,) but the fact that they do
not display a history-contingent\cite{Gell-Mann-Hartle} or branch-dependent
character or, in other words, that they are {\it repeatable}.
Furthermore,``repeatable'' does not mean that we can guarantee for each and
every dynamical variable to repeat its history every time we set up another
run of the experiment. What it actually means is we can separate in a
consistent and useful manner the dynamical variables we are not interested
in and average over when the dust has settled. In doing this, we must end up
with a mathematical construct that succeeds to statistically describe the
behaviour of the whole collectivity. Another aspect of experiments, as
opposed to more general situations, is they are limited by a series of
specific {\it consistency conditions} that must be put in and duly
justified. This point will be further developed later.

2$^{{\rm {nd}}}$: Although a description based on the density matrix is the
minimal guaranteed on theoretical grounds\cite{Gleason}, I assume that the
density matrix is to be obtained from a more fundamental description based
on state vectors (again, a view not universally agreed upon\cite{Ballentine}%
). The density-matrix description arises, then, due to a combination of two
things:

(i) Entanglement between the sub-system we are interested in and the rest of
the world (including, if necessary, distant systems\footnote{%
This is a reason (among others) to be extremely careful when ascribing the
occurrence of decoherence to an ``environment''. The part of the whole
system responsible for the leading role in the appearance of decoherence in
the ``interesting'' system can, in principle, be very far away.})

(ii) Ignorance (or unconcern) on our part about the fine details of the
latter's history.

This mechanism is sometimes described (see e.g., \cite{Zurek} for a brief
exposition) in the following terms: if a typical pure entangled state of the
whole system (${\cal S}$)$+$environment (${\cal E}$) is $\left| \Psi _{{\cal %
SE}}\right\rangle =\sum_qc_q\left| q,{\cal E}_q\right\rangle $, then writing
down the operator $W_{{\cal SE}}\equiv \left| \Psi _{{\cal SE}}\right\rangle
\left\langle \Psi _{{\cal SE}}\right| $ produces

\begin{equation}
\label{density-matrix1}W_{{\cal SE}}=\sum_{q,q^{\prime }}c_qc_{q^{\prime
}}^{*}\left| q,{\cal E}_q\right\rangle \left\langle q^{\prime },{\cal E}%
_{q^{\prime }}\right| 
\end{equation}
which, after summing over the ``uninteresting'' environmental degrees of
freedom ${\cal E}_q$ and assuming $\sum_q\left| {\cal E}_q\right\rangle
\left\langle {\cal E}_q\right| =I_{{\cal E}}$, yields

\begin{equation}
\label{density-matrix2}W_{{\cal S}}=\sum_q\left| c_q\right| ^2\left|
q\right\rangle \left\langle q\right| 
\end{equation}
while the most general $\left| \Psi _{{\cal SE}}\right\rangle $ displaying
entanglement between our system ($q$-variables) and the rest of the world ($%
{\cal E}$-variables) would lead to

\begin{equation}
\label{density-matrix3}W_{{\rm {SE}}}=\sum_{q,q^{\prime }}c_{q{\cal E}%
}c_{q^{\prime }{\cal E}^{\prime }}^{*}\left| q,{\cal E}\right\rangle
\left\langle q^{\prime },{\cal E}^{\prime }\right| 
\end{equation}

But, as is easy to check by using the closure relation for the ${\cal E}$'s:

\begin{equation}
\label{density-matrix4}W_{{\cal S}}=\sum_{{\cal E}}\left\langle {\cal E}%
\left| W_{{\cal SE}}\right| {\cal E}\right\rangle =\sum_{q,q^{\prime
}}W_{qq^{\prime }}\left| q\right\rangle \left\langle q^{\prime }\right| 
\end{equation}
where $W_{qq^{\prime }}=\sum_{{\cal E}}c_{q{\cal E}}c_{q^{\prime }{\cal E}%
}^{*}$. Conclusion: a more general pattern of entanglement between the
system and the rest of the world than the ``matching'' one in (\ref{density-matrix1})
 does {\it not} result in a decoherent expression for $W_{%
{\rm {S}}}$. Thus, while the punch-line of the previous argument is
essentially correct, the ambiguous word ``environment'' can render it
somewhat misleading. What is really essential to the argument is: a typical
decoherent situation is one in which some extraneous system has so special a
relation to our system of interest as to repeatedly interact with it through
a selected collection of states $\left| {\cal E}_q\right\rangle $ that
exactly replicates the system's (normal environment, ``apparatus'', or
whatever other technical name).

However, if the viewpoint of attaching a non fundamental meaning to $W$ is
to be embraced, it is of the greatest importance to take to heart the {\it %
merely instrumental nature of the density matrix}. This operator is only an
artifact I use in order to calculate probabilities relative to whatever
experiment I decide to carry out, and is not an attribute of individual
systems. It must be let for those who look at $W$ as a fundamental
description to lament, e. g., the impossibility to write it unambiguously in
terms of state vectors.

3$^{{\rm {rd}}}$: Formula (\ref{collapse}) is dismissed. Quantum evolution
(whatever it ultimately represents) is always linear and unitary; monitored
by the Schr\"odinger equation. I claim though, the soundness of making
approximations {\it on it,} which must be duly justified. Furthermore, I
will supplement dynamical quantum mechanics with the position of the
particle and its trajectory in order to establish a more detailed dynamical
theory.

The addition of the particle's position and trajectory will be realised here
according to Bohmian mechanics, which is maximally abridged: It branches
into two levels:

(BM1): Waves that evolve according to (\ref{Schrodinger-evolution}) with $%
U(t_2,t_1)$ a unitary linear operator being a function of a so-called CSCO
(Complete Set of Commuting Operators).

(BM2): Wave-bound particles; the corresponding trajectories given by%
\footnote{%
Straightforward manipulations lead to this form from the nowadays favoured
one $m{\bf v}({\bf x},t)=\hbar $Im$\left( {\bf \nabla }\psi ({\bf x},t)/\psi
({\bf x},t)\right) $ or $m{\bf v}({\bf x},t)=\hbar $Im$\left( \psi ^{*}({\bf %
x},t){\bf \nabla }\psi ({\bf x},t)/\psi ^{*}({\bf x},t)\psi ({\bf x}%
,t)\right) $ (see, e.g., \cite{Durr-Goldstein-Zanghi}). The latter is more
convenient as concerns arguments on probability flux and related matters.
But in the present discussion, I will need the proposed form (\ref{BM-trajectories})
 because it involves the logarithm. The continuous
occurrence of its complex-plane inverse; the exponential operator, in the
present work makes this form compelling.}

\begin{equation}
\label{BM-trajectories}m{\bf v}({\bf x},t)=\hbar {\bf \nabla }{\rm Im}\left(
\ln \psi ({\bf x},t)\right) 
\end{equation}
where $\psi ({\bf x},t)=\left\langle {\bf x}\left| \psi (t)\right.
\right\rangle $ is the unique solution of (\ref{Schrodinger-evolution})
furnished with the inicial conditions implementing the specific experimental
preparation and I assume $\arg \psi ({\bf x},t)\in \left[ 0,2\pi \right) $
to fix ideas, although none of the results depend on the particular Riemann
sheet we place the logarithm in (due to the action of the gradient).

But if Bohmian mechanics is to crown its aspirations to explain\footnote{%
Mind my use of the word ``explain,'' and not ``supersede''. Bohmian
mechanics cannot be dismissed on the grounds that it is incapable of getting
``the same results with as little effort'' as CQM. As long as we succeed to
prove CQM to be but a user-friendly version of the more fundamental Bohmian
mechanics, the effort spent in using Bohmian mechanics to solve ``typically
quantum'' problems can be compared to the effort spent in using general
relativity to calculate the orbits of a free-falling marble. Though, it is
certainly uncomfortable that no context has appeared as yet in which Bohmian
mechanics goes {\it further} than CQM in the phenomenological battleground.}
CQM's famous FAPP validity, some demonstration is needed that the
identification of squared amplitudes with probabilities is a sound
procedure. This would imply, either a mathematical demonstration, (probably
based on an analysis of Bohmian evolution {\it \`a la Liouville}) that
Bohm's evolution leads to such a quantum equilibrium situation from
arbitrary initial conditions, or else a more challenging step-by-step
argument involving a discussion of every conceivable experimental {\it %
preparation} (they are not as many as its seems at first sight,) e.g.,
collimation, thermal bath at finite temperature, etc. by using simplifying
auxiliary hypotheses specific to each example. This is an ongoing program of
research in Bohmian mechanics that I will not deal with in the present work.
I am just going to draw conclusions from the assumption of compliance with
the quantum equilibrium condition.

But I will have to appeal to the reader's patience and engage (once more!)
in some further excruciating use of trite quantum formalism. Though not
earth-shattering by any means, there are some new insights in store for
patient readers.

\section{Brief Mathematical Notices.}

My reasoning will very much rely on overlooked consequences of well-known
mathematical facts, so it will be convenient to summarise these facts.
Remember: provided a space of accessible states is specified, we can expand
any self-adjoint operator $F$, such that $\left[ F,Q\right] =0$ as

\begin{equation}
\label{operator-expansion1}F=\sum_{q\in \sigma (Q),\lambda }f(q,\lambda
)P_{q,\lambda } 
\end{equation}
with $P_{q,\lambda }=\left| q,\lambda \right\rangle \left\langle q,\lambda
\right| $ and some index $\lambda $ accounting for degeneracy.

Several remarks to be made: (1) An offensively redundant wording of (\ref{operator-expansion1})
: the fact that $F$ is $Q$-commuting plainly means
that $F$ is a linear combination of $q$-projectors (the point is essential).
(2) The summatory symbol could mean a Lebesgue integral for some sub-regions
of the spectrum $\sigma (Q)$ and I will shift from one notation to the other
somewhat relaxedly. (3) Not every operator acting on the space of states
complies with (\ref{operator-expansion1}), because an arbitrary operator
could depend on further operators, not all of them commuting with $Q$. The
extended set which does constitute a real operator basis is called an ISO
(Irreducible Set of Observables).

Result (\ref{operator-expansion1}) entitles us to the notation $f(Q)$. A
so-called operator calculus based on this relation is possible. Dropping $%
\lambda $ from the equation for the sake of simplicity, and keeping it
always in mind as an eventual addendum:

\begin{equation}
\label{operator-expansion2}f(Q)=\sum_{q\in \sigma (Q)}f(q)P_q 
\end{equation}
of which $Q$ itself is a particular instance $Q=\sum_{q\in \sigma (Q)}qP_q$.

Further technical questions need not be detailed here. What I do need is to
prompt a series of immediate corollaries of (\ref{operator-expansion2}) for
particular instances it contemplates. First, if $f(Q)=\exp iQ$

\begin{equation}
\label{exponential}f(Q)=\exp iQ=\exp \left( \sum_{q\in \sigma
(Q)}iqP_q\right) =\sum_{q\in \sigma (Q)}e^{iq}P_q 
\end{equation}

A direct application of (\ref{exponential}) is provided by the action of $%
U(t_2,t_1)$ on an arbitrary state when the latter is expanded in eigenstates
of a Hamiltonian with a bounding potential. Then we have: $Q=-(t_2-t_1)H_{%
{\rm {bounding}}}/\hbar $ and $q$ is simply $-(t_2-t_1)/\hbar $ times the
energy of the corresponding bound state.

But one could foster hopes for the profitability of (\ref{exponential}) to
reach further than that, for it is the exponential operator of $%
-i(t_2-t_1)H/\hbar $ which, in the most general context, carries quantum
states from time $t_1$ to $t_2$. Furthermore, (\ref{exponential}) involves
two among the most relevant elements in quantum theory of measurement; one
is projectors, and the other is phase factors (which must have {\it something%
} to do with how decoherence comes about). Of course, in a general case I
cannot take advantage of (\ref{exponential}), because the dynamics is far
too messy in terms of the operator $Q$ I may be interested in (i.e., $H$
does not commute with it). The whole idea of the present work is to show
that actually, there are instances in which we {\it can} do such a
simplification. In fact, as we will see, I can define an ``ideal''
measurement as a situation in which, at some crucial time, the dynamics
undergoes a transitory stage characterised by (\ref{exponential})\footnote{%
To be sure, an ideal measurement {\it is} implicitly defined in this way in
the literature. See e.g., an exhaustively documented reference such as \cite{Galindo-Pascual}
 or the classics \cite{Bell-Nauenberg} or \cite{Wheeler-Zurek}
 in this concern. There you can find an ideal measurement
characterised as one in which neither (1$^{{\rm {st}}}$): the weights $%
w_q=\sum_\lambda \left\langle q,\lambda \left| w\right| q,\lambda
\right\rangle $ of the different eigenstates nor (2$^{{\rm {nd}}}$): the
coherences $w_{\lambda \lambda ^{,}}=\sum_q\left\langle q,\lambda \left|
w\right| q,\lambda ^{,}\right\rangle $ between $Q$-commuting variables are
altered after the measurement process. This statement is equivalent to
statement (\ref{diagonal-H}) that we will see next (theorem). And (\ref{exponential})
 is a further corollary resulting when the evolution formula
is considered. The 2$^{{\rm {nd}}}$ requirement amounts to requiring that
the evolution operator factorises into a non-trivial part acting on the
subspace spanned by $Q$-eigenstates times the identity operator acting on
the subspace spanned by $\Lambda $-eigenstates, where $\lambda $ is a
generic eigenvalue corresponding to a certain operator $\Lambda $.}.

I introduce also the following device: I do not demand that the collection $%
\left\{ P_q\right\} $ be {\it exhaustive}\footnote{%
Some authors call exhaustive any collection of projectors that sum up to the
identity operator $\sum_kP_k=I$. We do not follow this use because it is
more correct to call such sets complete (as is otherwise traditional,) and
we do need to distinguish them from the really exhaustive ones. For us, $%
\left\{ P_q\right\} $ is exhaustive if {\it any} operator $f(Q)$ admits an
expansion: $f(Q)=\sum_{q\in \sigma (Q)}f(q)P_q$. This is in keeping with the
difference in nuance between both words in ordinary English.}, that is; even
if $\sigma (Q)$ is continuous, I can perform a coarse-graining by
integrating the point-like projectors over certain extended regions $D_k$%
\footnote{%
Our coarse graining is meant to be relevant at a specific time. In
particular, we will not use it to build a sequential product of operators or 
{\it history} for the observable in question. In this respect (and some
more,) this approach differs from the loosely equivalent ones taken up by
Griffiths, Gell-Mann-Hartle\cite{Gell-Mann-Hartle}, Omn\'es\cite{Omnes},
Zurek\cite{Zurek}, etc.}. Then

\begin{equation}
\label{coarse-graining1}P_k(Q)\equiv \int_{D_k}dqP_q 
\end{equation}
for a numerable partition $\left\{ D_k\right\} _{k=1,\cdots n}$ of $\sigma
(Q)$. Nothing prevents me from adopting a partition with an infinite number
of projectors, but I keep it finite for reasons that will become apparent
later. This finiteness will presumably force upon me to take some of the $%
D_k $ infinite in extension (again, a feature that will have a suitable
explanation). Then, if $\alpha _k$ are real numbers, and if I previously
introduce

\begin{equation}
\label{coarse-graining2}g(Q)=\sum_{k=1}^ni\alpha _kP_k(Q) 
\end{equation}
the application of (\ref{operator-expansion2}) to $f=\exp \circ g$ renders
formula (\ref{exponential}) still validated as a new ``coarse-grained''
version

\begin{equation}
\label{coarse-grained-exponential}\exp \left( i\sum_{k=1}^n\alpha
_kP_k(Q)\right) =\sum_{k=1}^ne^{i\alpha _k}P_k(Q) 
\end{equation}

Mind the purposeful nature of the discrete index $k$. Its occurrence means
that $f(Q)$ in this operator calculus is now a function of $Q$ {\it through}
the coarse-graining just introduced, and not an arbitrary function of $Q$.
Any doubts that (\ref{coarse-grained-exponential}) is valid after such a
procedure, are dissipated by directly expanding the power series plus use of
orthogonality relations $P_k(Q)P_j(Q)=\delta _{kj}P_j(Q)$.

Another corollary of (\ref{exponential}) that will prove useful: let us
expand $Q$ as $Q=R\otimes S$ for certain complete operator set $R$ and $S$.
The reader does not need a boring proof to believe

\begin{equation}
\label{2-factor-exponential}\exp \left( i\sum_{r,s}\alpha _{rs}P_r\otimes
P_s\right) =\sum_{r,s}e^{i\alpha _{rs}}P_r\otimes P_s 
\end{equation}
but if suspicious again is invited to check it the hard way by direct
expansion and using along the way identities as $\left( P_1\otimes
P_2\right) ^{\perp }=P_1^{\perp }\otimes P_2^{\perp }+P_1\otimes P_2^{\perp
}+P_1^{\perp }\otimes P_2$. Be careful with formula (\ref{2-factor-exponential})
: I have found fairly seasoned postgraduates in
theoretical physics taking offence at my emphasising it (because of its
triviality\footnote{%
And trivial {\it is is not} in several ways: the reader may well be used to
its common occurrence in matrix algebra. What is almost universally ignored
is the validity of this kind of coarse-grained-version identities that makes
them illuminating when applied within the position factor of the overall
Hilbert space.}) that actually do fail to apply it correctly right away. For
example, do not rush into:

\begin{equation}
\exp \left( i\alpha P_1\otimes P_2\right) =e^{i\alpha }P_1\otimes P_2{\rm {\
\;\;\;(false)}} 
\end{equation}
for $P_1+P_2=I$, but mind instead that

\begin{equation}
\exp \left( i\alpha P_1\otimes P_2\right) =e^{i\alpha }P_1\otimes
P_2+P_1^{\perp }\otimes P_2^{\perp }+P_1\otimes P_2^{\perp }+P_1^{\perp
}\otimes P_2{\rm {\;\;\;(true)}} 
\end{equation}
because any $\alpha _{rs}=0$ in the l.h.s. of (\ref{2-factor-exponential})
transforms into $e^{i\alpha _{rs}}=1$ in the r.h.s. of (\ref{2-factor-exponential})
 (sorry for the lag to the cleverer people). Another
operator identity I will also use is

\begin{equation}
\label{tensor-exponential}\exp iQ\otimes R=\exp \left( \sum_{q\in \sigma
(Q)}iqP_q\otimes R\right) =\sum_{q\in \sigma (Q)}P_q\otimes \exp \left(
iqR\right) 
\end{equation}

Finally, a further relation I will need is the Campbell-Haussdorff identity
for two non-commuting operators $A$ and $B$

\begin{equation}
\label{Campbell-Haussdorff}\exp A\exp B=\exp \eta (A,B) 
\end{equation}
whose initial terms are

\begin{equation}
\label{eta-definition}\eta (A,B)=A+B+\frac 12\left[ A,B\right] +\frac
1{12}\left[ \left[ A,B\right] ,B\right] +\frac 1{12}\left[ \left[ B,A\right]
,A\right] +\cdots 
\end{equation}

Later, I will make use of (\ref{eta-definition}) in relation to the kinetic
and potential-energy operators. The commutator of these operators produces a
non constant, non-$A$ nor $B$-commuting operator in general, rendering (\ref{Campbell-Haussdorff})
 almost unmanageable. But an approximation will be
introduced and proved feasible for certain states upon which (\ref{Campbell-Haussdorff})
 will be made to act. The essence of such
approximation is that the action of (\ref{eta-definition}) upon such
suitable states (that is, for appropriate values of certain parameters),
yields successive commutators that can be made arbitrarily small, so I will
not need to care about fine details of this formula (numeric factors and
signs) as long as we keep in mind that every term is obtained by further
commutation of a previous-order term with either $A$ or $B$ and
multiplication by a numeric factor.

\section{Coarse-Grained Evolution Formulae and \penalty-10000 Bohm's Approach to
Measurement.}

I am now going to take a glimpse at history with the help of the previous
tools. In doing so, one realises that certain basic results of a
long-established analysis in the quantum theory of measurement\cite{Bohm},%
\cite{Wheeler-Zurek} are but a trivial application of operator formula (\ref{exponential})
 (or (\ref{coarse-grained-exponential}) when a coarse-graining
is required). I also notice that there is nothing to prevent me from
applying Bohm's analysis to the particular instance of a position
measurement. In fact, I can contemplate a localisation as a particular
example among a series of other paradigmatic examples all of them cast into
the same mathematical mold. But paying heed to the physics lore\cite{Bell-Nauenberg}
, a localisation is not just any old kind of measurement. It
is {\it the} measurement. That is why, in the present work, this classical
analysis by Bohm is taken as all but conclusive.

But let us shortly review Bohm's approach. It goes like this: if we are to
``measure'' the quantum variable $Q$, the dynamical approximation that is in
force at a certain critical time during that process (to be further
specified) is one that makes the Hamiltonian operator diagonal in $Q$%
\footnote{%
It is frequent to find an even more restrictive assumption in the
literature, such as $H_Q=Q\otimes P_y$ (the apparatus being included in the
analysis,) with $P_y$ being a canonical momentum associated to the
apparatus. This is an extension to a tensor-product space compatible with (%
\ref{diagonal-H}) that I will consider later in a slightly modified form.
For the time being, I am focusing on how things look from the one-factor
Hilbert space of the system that is being analised.}, that is, I consider
some

\begin{equation}
\label{diagonal-H}H_{Q{\rm {-measuring}}}=H\left( Q;Q{\rm {-{%
commuting\,operators}}}\right) 
\end{equation}

It is unfortunate that no mathematical notation can justly enfazise the
interesting features of the previous formula. The suffix in the l.h.s. of it
does {\it not} imply $Q$-dependence. It implies whatever technical
conditions the experimentalist has to comply with to make sure the
experiment produces the results it has to. The r.h.s. does speak of $Q$%
-dependence, so the relation is very far from obvious. In order to properly
understand how relation (\ref{diagonal-H}) comes about, it is far better to
drop any kind of aprioristic reasoning (or blind faith in a messianic
insight by its original proposer) and go instead one by one to the
particular examples we know best. Thus, I use an inductive reasoning to see
that (\ref{diagonal-H}) is in fact a good account of things going on at
least in several standard situations, as long as they are seen as a
limiting, well-behaved case\footnote{%
A further remark, though, is necessary: I must in this concern ignore
several technical qualifications about different kinds of measurement that
would obscure the point I am trying to bring to light, although they could
prove relevant to some other effects. I mean those referred to as QND
(quantum non demolition), 1$^{{\rm {st}}}$ and 2$^{{\rm {nd}}}$ kind
measurements, etc. Thus, a general process of measurement could involve
several stages, each sharply defined as concerns (\ref{diagonal-H}), and,
consequently, each equally suitable for its application. Yet, some of these
stages could stand for an example of QND etc., while others would not.}.

Consider, e.g., the Stern-Gerlach experiment. The first stage, in which a
particle is selected that approaches the Stern-Gerlach window in the desired
direction, constitutes a linear momentum measurement (preparation) in
itself, with a state evolving in accordance with the free-evolution formula:

\begin{equation}
\label{free-evolution}\left| \psi {\bf (}t_2)\right\rangle =e^{-i\eta
(t_2-t_1){\bf P}^2/2m\hbar }\left| \psi {\bf (}t_1)\right\rangle 
\end{equation}

This is a nice example of spectral formula (\ref{exponential}) as clearly
seen by expanding: 
\begin{equation}
\left| \psi {\bf (}t_2)\right\rangle =\int d^3p\left| {\bf p}\right\rangle
\left\langle {\bf p}\right| e^{-i\eta (t_2-t_1){\bf P}^2/2m\hbar }\left|
\psi {\bf (}t_1)\right\rangle = 
\end{equation}

\begin{equation}
\int d^3pe^{-i\eta (t_2-t_1){\bf p}^2/2m\hbar }\left| {\bf p}\right\rangle
\left\langle {\bf p}\right. \left| \psi {\bf (}t_1)\right\rangle =\sum_{{\bf %
p}}e^{-i\eta (t_2-t_1){\bf p}^2/2m\hbar }P_{{\bf p}}\left| \psi {\bf (}%
t_1)\right\rangle 
\end{equation}
and consequently, of the expression already advanced (\ref{multi-projection}%
). Being $H$ (hence $U(t_2,t_1)$) diagonal in ${\bf P}$, I end up recovering
the well-known free-evolution formula. Of course, you need not go this
length for me to teach you free-quantum evolution. The point to highlight is
(\ref{free-evolution}) as being a particular instance for the application of
(\ref{exponential}).

Suppose now that at $t=t_3$, free evolution has accomplished its job of
getting an incoming state fairly peaked in the momentum space and in the
desired direction. This is the time when the quantum wave reaches the
Stern-Gerlach window, at which moment an interaction impulsive in nature is
triggered within the window (while $\left[ t_1,t_2\right] $ is extremely
large, the impulsive interval $\left[ t_2,t_3\right] $ must be very short%
\footnote{%
``Large'' or ``short'' mean, of course, as compared to the relaxation time
provided by the dispersion of a wave packet. In order to be definite, I can
take $\Delta t=\Delta _\psi K({\bf P})/\left| \partial _t\left\langle K({\bf %
P})\right\rangle _\psi \right| $ in the spirit of the Mandelstam-Tamm
interpretation of the uncertainty relation\cite{Mandelstam-Tamm}.}). This
interaction is given account of by the magnetic-dipole term $H\propto \sigma
_nB_n\left( {\bf x}\cdot {\bf n}\right) $, so we have rule (\ref{diagonal-H}%
) again for the particular instance $Q=\sigma _n$. But take notice: the
experimenter does not set out to comply with condition (\ref{diagonal-H}) as
a procedural experimental prescription; the dipole Hamiltonian representing
the experimenter's manipulations is {\it inevitably} diagonal in $\sigma _n$.

Subsequently, a time lapse from $t_3$ to a certain $t_4$ is necessary for
the beam to reach a faraway screen. This last stage of the experiment
requires (sticking to what we think is the surest commandment of quantum
mechanical formalism, i.e., the Schr\"odinger equation!) an interaction that
ideally, would look very much like $H\simeq V({\bf x})$ (collision). But
then rule (\ref{diagonal-H}) is in force once again for the election $Q={\bf %
X}$.

Some further elaboration of the previous example showing how (\ref{diagonal-H})
 cannot be too far off the mark in matters of measurement is
given next, but let us recall first as the last preliminary example that an
energy {\it preparation} for an stationary state involves a dynamics
monitored by a Hamiltonian that is diagonal (as couldn't be otherwise) in
its own representation: $H=\sum_EEP_E\Longrightarrow \exp \left[
-i(t_2-t_1)H/\hbar \right] =e^{-i(t_2-t_1)E/\hbar }P_E$.

Thus, completely different experimental procedures coincide in mathematical
form under a conceptual unification that involves both preparations and
other, of more uncontrolable effects, measurements.

\subsection{The Stern-Gerlach experiment.}

Here I fill up some details of this already outlined most classical piece of
theoretical analysis about measurement with the fresh feature of showing
that it is simply a particular instance of identity (\ref{exponential}). Let
us take the spin-1/2 Hilbert space for neutral\footnote{%
We put off inconvenient dragging Lorentz forces.} paramagnetic particles,
and $P_{+}=\left| +\right\rangle \left\langle +\right| $, $P_{-}=\left|
-\right\rangle \left\langle -\right| $. The $z$-component of spin is
represented by the operator $\sigma _z=P_{+}-P_{-}$. Given that the
Hamiltonian interaction at the passing of the beam through the Stern-Gerlach
window can be approximated by $H=-\mu _z\sigma _zB_z(z)$, after application
of (\ref{exponential}) to the particular case of $U(t_2+\tau ,0)$ with $t_2$
being the time when the wave packet enters the window, and assuming an
initial state approaching the window along the $x$-axis: $\left\langle {\bf x%
}\mid \psi (t_2)\right\rangle =e^{ip_0x/\hbar }\left( \psi _{+}({\bf x}%
)\left| +\right\rangle +\psi _{-}({\bf x})\left| -\right\rangle \right) $, I
get

\begin{equation}
\label{1}\left\langle {\bf x}\mid \Psi (t_2+\tau )\right\rangle
=e^{ip_0x/\hbar }\left( e^{i\alpha }e^{iz\Delta p_z/\hbar }\psi _{+}({\bf x}%
)\left| +\right\rangle +e^{-i\alpha }e^{-iz\Delta p_z/\hbar }\psi _{-}({\bf x%
})\left| -\right\rangle \right) 
\end{equation}

The deflection is given by both factors $e^{\pm iz\Delta p_z/\hbar }$ coming
from Taylor-expanding the magnetic field inside the Stern-Gerlach window,
with $\Delta p_z=\mu _z\partial B_z(z)/\partial z$ evaluated at $x=y=z=0$
(the centre of the window). The remaining parameter $\alpha =\mu _zB_z(0)$
and other analogous appearing later will be interpreted (as Bohm\cite{Bohm}
already did without still considering additional variables,) as giving rise
to decoherence when Bohmian mechanics is added to the scheme. Bohmian
mechanics postulates as a plausible guess (to be further consolidated
theoretically) that the quantum equilibrium condition is universally in
force. This implies that the wave function's modulus squared be used as a
probability density for the {\it localisation} of the individual particles.
But if such wave function is affected by further (stochastic) parameters,
the next reasonable step to follow is to introduce some average over these
parameters in addition to the quantum scalar product. At first, I will take
this average quite relaxedly, simply justiftying it upon the fact that such
parameters generally appear as phase factors varying very rapidly in
relation to all the remaining quantum parameters involved. Later, I will
involve myself more deeply with this question and formally introduce such an
average after assuming entanglement between the system under study and the
apparatus. The mentioned additional average will contemplate, among other
things, a Hilbert scalar product for the pointer variables of the apparatus.
As a further historical remark, let us say that in none of Bohm's works is
there an explicit reference to, nor any use of identity (\ref{exponential}).

\subsection{The special role of position measurements.}

The next one is an example that really provides telltale clues in matters of
interpretation and is a simple corollary of the coarse-grained version (\ref{coarse-grained-exponential})
 of our identity. The reason why Bohm did not
(to the best of my knowledge) pay attention to the possibility of this
argument is to realise that, back then, it was not very fashionable to look
at quantum mechanics in the light of V. Neumann's spectral analysis of
yes-no observables, as it has been, e.g., in the sequel of works \cite{Griffiths}
, \cite{Gell-Mann-Hartle}, \cite{Omnes}, etc.

To develop the argument, I use the impulsive approximation\footnote{%
Strictly speaking, the impulsive approximation not only implies the
aforementioned ${\bf X}$-dependence, but {\it at the same time,} the time
dependence $\chi _{\left[ t_1,t_2\right] }(t)$ implementing its brief
validity in time. It is, of course, both possible and sterile here to be
somewhat more formal and use Dyson's time-dependent evolution formula $\exp
\left[ -i/\hbar {\cal P}\int H(t)dt\right] $ with $H(t)=\chi _{\left[
t_1,t_2\right] }(t)V({\bf X})$.} $H\simeq V({\bf X})$. Furthermore, I pick
the particular realisation of (\ref{coarse-grained-exponential}) with $Q=%
{\bf X}$, and adopt a convenient coarse graining $P_k({\bf X})\equiv
\int_{D_k}d^3x\left| {\bf x}\right\rangle \left\langle {\bf x}\right| $
whose justification ultimately must be put to rest on the existence (whether
on purpose or not) of a certain finite set of non-overlapping regions $D_k$
where the particles can be captured (``detectors'', you may think). This
produces in position representation

\begin{equation}
\label{2}\left\langle {\bf x}\left| P_k({\bf X})\right| {\bf x}^{\prime
}\right\rangle =\chi _{D_k}({\bf x})\delta ({\bf x-x}^{\prime }) 
\end{equation}

But then, with the substitution $\alpha _k=-\eta _k(t_2-t_1)/\hbar $:

\begin{equation}
\label{3}\left\langle {\bf x}\left| \exp \left( i\sum_{k=1}^n\alpha _kP_k(%
{\bf X})\right) \right| {\bf x}^{\prime }\right\rangle =\delta ({\bf x-x}%
^{\prime })\sum_{k=1}^ne^{-i\eta _k(t_2-t_1)/\hbar }\chi _{D_k}({\bf x}) 
\end{equation}
so that:

\begin{equation}
\label{4}\psi ({\bf x,}t_2)=\sum_{k=1}^\infty e^{-i\eta _k(t_2-t_1)/\hbar
}\chi _{D_k}({\bf x})\psi ({\bf x,}t_1) 
\end{equation}

To be more concrete, imagine now the particular realisation of (\ref{4})
that models the interaction at a certain moment $t_1$ is precisely a sharp
potential ``basin'' $V({\bf x})=\eta \chi _D({\bf x})$, $\eta <0$ or a
``plateau'' $\eta >0$ where $D$ is some particular $D_k$ (a rough modelling
of a definite ``detection''). I would have

\begin{equation}
\label{5}\psi ({\bf x,}t_2)=e^{-i\eta (t_2-t_1)/\hbar }\chi _D({\bf x})\psi (%
{\bf x,}t_1)+\chi _D^{\perp }({\bf x})\psi ({\bf x,}t_1) 
\end{equation}

Now eq. (\ref{5}) can be qualified anything we want except neutral in
matters of interpretation: it is sure to give a hard time to anyone
asserting CQM is complete and willing to consider it seriously, and make the
delight of a ``hidden-variable'' advocate. It simply is telling us that the
wave function, when suddenly ``hit'' by an impulsive localised square
potential does not change its associated local probability density $\left|
\psi ({\bf x,}t)\right| ^2$ at all!. It just shifts the relative phase
between the inside component $\psi _{{\rm {in}}}({\bf x})\equiv \chi _D({\bf %
x})\psi ({\bf x})$ and the outside one $\psi _{{\rm {ex}}}({\bf x})\equiv
\chi _D^{\perp }({\bf x})\psi ({\bf x})$ (self-explanatory definitions). It
is, besides, unable by itself to provide an image of what a detection must
be. But if someone is to dismiss the equation as nonsense, they should also
explain why so unmistakable an approximation as this (no doubt that $%
H(t)=\chi _{\left[ t_1,t_2\right] }(t)V({\bf X})$ does represent however
crudely a localising attempt) fails so catastrophically to embody the
common-sense properties of localisation. I am going to take delight in it
and show how, when combined with the general scheme of Bohmian mechanics, it
explains the appearance and persistence of decoherence for times $t$
corresponding to the arrival of the wave fronts at a faraway screen or
detector. Moreover, I am going to show that, in combination with Bohmian
mechanics, eq. (\ref{5}) provides an image of particles bouncing off $D$ by
either absorbing ($\eta <0$) or repelling ($\eta >0$) the particle with an
uncertain direction depending both on $D$ and the state's spacial profile.
Finally, I am also going to get an unmistakable picture of quantum EPR
``nonlocality'' in the last section of this work.

With a more general potential, the plain impulsive approximation giving $%
\psi ({\bf x,}t_2)=e^{-iV({\bf x})(t_2-t_1)/\hbar }\psi ({\bf x,}t_1)$,
imprints a locally varying phase shift in the outgoing wave function or, in
other words, changes its momentum distribution in a way easily analysed by
applying the momentum operator to the outcoming wave. If $\tau =t_2-t_1$,
the momentum distribution at $t=t_2$ is given by

\begin{equation}
\left[ {\bf P}\psi \right] ({\bf x,}t_2)=-i\hbar {\bf \nabla }\left[ e^{-iV(%
{\bf x})\tau /\hbar }\psi ({\bf x,}t_1)\right] = 
\end{equation}

\begin{equation}
\label{6}\left[ -\tau {\bf \nabla }V({\bf x})\psi ({\bf x,}t_1)-i\hbar {\bf %
\nabla }\psi ({\bf x,}t_1)\right] e^{-iV({\bf x})\tau /\hbar } 
\end{equation}

No mistake about the interpretation of this: the new momentum distribution
is such that

\begin{equation}
\left\langle \psi (t_2)\right| {\bf P}\left| \psi (t_2)\right\rangle =\int
dx\psi ^{*}({\bf x,}t_1)\left[ -\tau \nabla V({\bf x})\psi ({\bf x,}%
t_1)-i\hbar \nabla \psi ({\bf x,}t_1)\right] = 
\end{equation}

\begin{equation}
\label{7}=\left\langle \psi (t_1)\right| {\bf P}\left| \psi
(t_1)\right\rangle -\tau \int dx{\bf \nabla }V({\bf x})\left| \psi ({\bf x,}%
t_1)\right| ^2 
\end{equation}

Now, this comes from commonplace Hamiltonian dynamics and is not mysterious
at all. The obvious interpretation of (\ref{7}) in terms of probability
distributions is: the wave function gets an additional ``kick'' in a
direction that is generally uncertain as long as the exact profile of $V(%
{\bf x})$ is uncertain, depending on both the potential and the initial
state's spatial profile.

On the other hand, in terms of Bohmian mechanics we get, by means of (\ref{BM-trajectories})

\begin{equation}
m{\bf v}({\bf x},t_2)=\frac \hbar {2i}\frac{\psi ^{*}({\bf x,}t_1){\bf %
\nabla }\psi ({\bf x,}t_1)-\psi ({\bf x,}t_1){\bf \nabla }\psi ^{*}({\bf x,}%
t_1)}{\psi ^{*}({\bf x,}t_1)\psi ({\bf x,}t_1)}-\tau {\bf \nabla }V({\bf x}%
)= 
\end{equation}

\begin{equation}
\label{7a}m{\bf v}({\bf x},t_1)-\tau {\bf \nabla }V({\bf x}) 
\end{equation}

so the new linear momentum is the previous one (given by the Bohmian
velocity field) plus a contribution $-\tau {\bf \nabla }V({\bf x})$. But
this is not the way in which the approximation proves itself more useful
(and plausible) to our purposes. The step previously considered of $V({\bf x}%
)$ assuming a square spatial profile with the stochastic character borne by
multiplying constants will prove itself better in a series of concerns as
will be shown in what follows.

To fix ideas (and also because of its paradigmatic character,) I consider
again a Stern-Gerlach experiment for a neutral particle with arbitrary spin $%
s$ assuming $2s+1$ possible values. First of all, I will reshape the general
set of projector-observables introduced before to render them tailor-made to
suit my experiment (here is the explanation of our previous
coarse-graining). So they are now a finite series of yes-no observables $P_m(%
{\bf X}),$ $m=-s,\cdots ,+s$, in the way of (\ref{2}), corresponding to $%
2s+1 $ non-overlapping ``detectors'', plus the complementary of the union $%
\cup _{k=-s}^{+s}D_m$, that is, $\left( \cup _{k=-s}^{+s}D_m\right) ^{{\rm {c%
}}}$ which will be denoted with the suffix ``ex'' for short. Each of these
``detectors'' corresponds to a region where the collision of a particle with
precisely that spin $z$-projection is expected to hit. Then we have a
partition of the identity in the position factor subspace of the whole
Hilbert space given by

\begin{equation}
\label{8}\sum_{m=-s}^{+s}P_{_m}+P_{{\rm {ex}}}=I 
\end{equation}

These tool-box operators constitute a drastic reduction as compared to the
more general formulation in quantum mechanics as concerns the implementation
of results by means of eigenvalues. It is a way of giving mathematical shape
to the feature that the only eigenvalues truly relevant to this experiment
when it comes to speaking of position measurements are certain, fixedly
defined {\it position q-bits}. So, it is two essential elements we are
integrating here; (1): impulsive interactions (dynamics) and (2): countable
character in the possible outcomes.

As an example, the particular operator corresponding to the
observable-proposition ``${\bf x}$ IS EXTERNAL TO $D_{-s}$'' would be
represented by the operator $P_{-s}^{\perp }=P_{-s+1}+P_{-s+2}+\cdots
+P_s+P_{{\rm {ex}}}$, and {\it its action on an eigenstate} $\left| \psi
\right\rangle $ {\it of the projectors making up} (\ref{8}) would be

\begin{equation}
\label{9}P_{-s}^{\perp }\left| \psi \right\rangle \equiv \left(
P_{-s+1}+P_{-s+2}+\cdots +P_s+P_{{\rm ex}}\right) \left| \psi \right\rangle
=\left\{ 
\begin{array}{c}
0\left| \psi \right\rangle 
{\rm \,if\,}\left| \psi \right\rangle {\rm \,is\,within\,}D_{-s} \\ 1\left|
\psi \right\rangle {\rm \,if\,}\left| \psi \right\rangle {\rm \,is\,outside\,%
}D_{-s} 
\end{array}
\right. 
\end{equation}

I can even consider, in principle, observables like: ``${\bf x}$ IS INTERNAL
TO EITHER $D_1$ OR $D_2$'', i.e.:

\begin{equation}
P_{\left( -s{\rm \,or\,}-s+1\right) }\left| \psi \right\rangle \equiv \left(
P_{-s}+P_{-s+1}\right) \left| \psi \right\rangle = 
\end{equation}

\begin{equation}
\left\{ 
\begin{array}{c}
1\left| \psi \right\rangle 
{\rm \,if\,}\left| \psi \right\rangle {\,{\rm \,is\,either\,within\,}}D_{-s}%
{\rm \,or}D_{-s+1} \\ 0\left| \psi \right\rangle {\rm \,if\,}\left| \psi
\right\rangle {\rm \,is\,outside\,both\,}D_{-s}{\rm \,and\,}D_{-s+1} 
\end{array}
\right. 
\end{equation}

Still other observables of classically impossible realisation, like ``${\bf x%
}$ IS INTERNAL TO BOTH $D_m$ AND $D_{m^{\prime }}$'' with $m\neq m^{\prime }$%
, which is identically assigned the zero operator.

But example (\ref{9}) leads us to the following: for this scheme to make any
sense within the context of my particular experiment, I need to supplement
this set of relevant projectors with a consistency condition on the states I
am using as possible inputs and outputs. In physical terms, this means I
have to aim my wave packets to the regions of detection and intently
preclude any situation in which any of these packets would be, at the moment
of detection ``caught in between'' $D_m$ and $\left( D_m\right) ^{{\rm {c}}}$
for arbitrary $m$. This means that the incoming wave is made up as a
coherent superposition of states that are either {\it completely within} or
else {\it completely without} the detection regions $D_m$ at the time
designed for each wave packet to hit ``its'' detector. In other words, I
will consider only states satisfying the {\it coincidence condition}:

\begin{equation}
\label{10}P_m({\bf X})\left| \psi _m(t_{{\rm {loc}}})\right\rangle =\delta
_{mm^{\prime }}\left| \psi _{m^{\prime }}(t_{{\rm {loc}}})\right\rangle 
\end{equation}

\begin{equation}
\label{11}P_{{\rm {ex}}}({\bf X})\left| \psi _m(t_{{\rm {loc}}%
})\right\rangle =0{\rm {\,for\,all\,}}m 
\end{equation}
{{\rm {\thinspace }}}with

\begin{equation}
\label{12}\left| \psi (t_{{\rm {loc}}})\right\rangle =\sum_{m=1}^nc_m\left|
\psi _m(t_{{\rm {loc}}})\right\rangle 
\end{equation}
where $t_{{\rm {loc}}}$ stands for the time when the localisation is
triggered. Mind the time dependence implied in (\ref{10})-(\ref{12}),
because I am focusing on experiments in which the initial state is prepared
as a superposition of wave packets (Gaussian, to be more concrete). Thus

\begin{equation}
\left\langle \left. {\bf x}\right| \psi (t)\right\rangle
=\sum_{k=-s}^{+s}c_m\left\langle \left. {\bf x}\right| \psi
_m(t)\right\rangle = 
\end{equation}

\begin{equation}
\label{13}\sum_{m=-s}^{+s}c_m\left( \sigma _m(t)\right) ^{-1/2}(2\pi
)^{-1/4}e^{i\alpha _m}e^{-i{\bf p}_m\cdot {\bf x}/\hbar }e^{-({\bf x}-{\bf x}%
_m(t))^2/4\sigma _m^2(t)} 
\end{equation}
where the parameters ${\bf p}_m={\bf p}_0+\Delta {\bf p}_m$, $\alpha _m=\mu
_mB_z(0)$ and $\Delta p_m=\mu _m\partial B_z(z)/\partial z$ come from the
previous stage of deflection as in the 2-state Stern-Gerlach experiment
reviewed before. Of course, (\ref{10})-(\ref{12}) do not determine $\left|
\psi _m(t_{{\rm {loc}}})\right\rangle $, so these are not eigenstates to be
uniquely determined (corresponding to the fact that the set of localisation
operators $P_m({\bf X})$ is not exhaustive).

A still more restrictive condition ({\it strong coincidence condition})%
\footnote{%
Great caution is needed, of course, to handle this condition in order not to
be led to inconsistencies, e.g.; the {\it strict} vanishing of the
derivatives of the wave function to every order {\it plus} the requisite on
the wave function to be analytic in ${\bf x}$, would lead for $\psi _m({\bf x%
},t){\bf \ }$to be identically zero. Reference \cite{Holstein-Swift}, is
given as a cautionary note to be taken in this concern. I am using the
conclussion of that analysis throughout; namely: the safest way to handle
states whose derivatives are being assumed to vanish to every order in a
certain region is by means of the propagator:} to be used in what follows,
is (\ref{10})-(\ref{12}) plus

\begin{equation}
\label{strong-coincidence}\left. \frac{\partial ^{\left( r\right) }}{{\left(
\partial x_k\right) }^r}\psi _m({\bf x,}t_{{\rm {loc}}})\right| _{\partial
D_m}=0{\rm {\,for\,all\,}}r 
\end{equation}
Although at first sight very strong a condition in a general setting, it is
always possible to make it valid by making the regions $D_m$ large enough.
We can play quite freely with such parameters because of the non existence
of a fundamental length parameter imposed upon us, the point of interest
being to discuss a plausible scenario for macroscopic localisation.

Let us digress a little about the previous ideas: a quite disturbing problem
the impulsive approximation suffers from is that it does not seem to embody 
{\it by itself} a macroscopic situation generally{\it \ }enough. The
question whether a more genuinely macroscopic condition can be used is given
a drastically simplifying answer by the model of{\it \ discontinuity in the
potential energy}. This model, though extremely idealised in the
mathematical side, when combined with the impulsive approximation seems a
more credible candidate for such macroscopicity condition than the bare
impulsive approximation. The following is suggested as a possible reason for
this: It is a well-known argument in statistical mechanics that for
non-analytic functions to occur as probability distributions (which is what
characterises the coexistence of distinct thermodynamic phases,) the
intervention of an infinite number of microscopic mechanical states is
strictly required. This is usually referred to as the problem of the
thermodynamic limit. If, in an analogous way, the occurrence in quantum
mechanics of sharp-edged potentials like $V({\bf x})=\sum_k\eta _k\chi _k(%
{\bf x})$ should come about because of the participation of very many
particles each with infinitely many states interacting with my particle of
interest, the previous model would constitute an intuitive {\it ex post facto%
} implementation of macroscopicity. In short, macroscopicity as concerns
localisation would be embodied by the ocurrence of discontinuous domains in
the potential energy. But this discontinuity-macroscopicity correspondence
should not be taken too far, because discontinuity in the potential energy
is also a feature of typically non-macroscopic problems as, e.g., reflection
on a potential wall. The coincidence condition plays a crucial role in that
it implements the fact that it is presumably the passing of the particle by
the region of detection what triggers the activation of such discontinuity
domains. In contrast, reflection problems (a situation which is widely used
in quantum experiments with no decoherence implied in it) require the
assumption that such discontinuity domains are pre-existing in the
particle's programmed path.

To show how the idea just exposed fits in without strictly having to make $K(%
{\bf p)}$ go to zero as compared to $V({\bf x})$ (which is what the usual
form of the impulsive approximation would require,) I make use of (\ref{Campbell-Haussdorff})
 and (\ref{eta-definition}), with $A=i\tau K({\bf P})$%
, $B=-i\tau \left( K({\bf P})+V({\bf X})\right) $, choosing $\hbar =1$ and $%
t_2-t_1\equiv \tau $:

\begin{equation}
\exp i\tau K\exp -i\tau \left( K+V\right) =\exp \eta \left( i\tau K,-i\tau
(K+V)\right) = 
\end{equation}

\begin{equation}
\exp \eta \left( i\tau K,-i\tau (K+V)\right) =\exp \left( -i\tau V+\frac{%
\tau ^2}2\left[ K,V\right] +\right. 
\end{equation}

\begin{equation}
\left. \frac{(i\tau )^3}6\left[ \left[ K,V\right] ,K\right] -\frac{(i\tau )^3%
}{12}\left[ \left[ K,V\right] ,V\right] +\cdots \right) 
\end{equation}

But notice that, when $\left\langle {\bf x}\left| K({\bf P})\right| {\bf x}%
^{\prime }\right\rangle =-(1/2m)(-i{\bf \nabla })^2\delta ({\bf x}-{\bf x}%
^{\prime })$, then

\begin{equation}
\left[ K(-i{\bf \nabla }),V({\bf x})\right] \psi \propto (-i{\bf \nabla }%
)^2(V({\bf x})\psi )-V({\bf x})(-i{\bf \nabla })^2\psi = 
\end{equation}

\begin{equation}
-\nabla ^2V({\bf x})\psi -2{\bf \nabla }V({\bf x})\cdot {\bf \nabla }\psi 
\end{equation}

The following terms are

\begin{equation}
\left[ \left[ K(-i{\bf \nabla }),V({\bf x})\right] ,K(-i{\bf \nabla }%
)\right] \psi \propto 
\end{equation}

\begin{equation}
=2{\bf \nabla }V({\bf x})\cdot {\bf \nabla }\left( \nabla ^2\psi \right)
-\nabla ^2(\nabla ^2V({\bf x}))\psi -4{\bf \nabla }\left( \nabla ^2V({\bf x}%
)\right) \cdot {\bf \nabla }\psi - 
\end{equation}

\begin{equation}
4(\partial _i\partial _jV({\bf x}))(\partial _i\partial _j\psi ) 
\end{equation}
and another one for $\left[ \left[ K(-i{\bf \nabla }),V({\bf x})\right] ,V(%
{\bf x})\right] \psi $, etc. The moral of the former expansion is: the
successive commutations are simply a sum of terms proportional to
derivatives of $V({\bf x})$ of order $\geq 1$, because the effect of
commutation is to remove the zero-order derivative of $V({\bf x})$ from $%
L\left[ V(x)\right] $ with $L$ being any polynomial differential operator
with constant coefficients. Now, if the class of states we are considering
is a superposition of say, wave packets $e^{-({\bf x}-{\bf x}_m(t_{{\rm {loc}%
}}))^2/4\sigma _m^2(t_{{\rm {loc}}})}$\footnote{%
I need to use functions that are suitable for the application of delta
distributions (that is; polynomically bounded).} satisfying the strong
coincidence condition at $t=t_{{\rm {loc}}}$ (negligibly small derivatives
at the boundary,) with $V({\bf x})=\sum_k\eta _k\chi _{D_k}({\bf x})$, it is
immediate that such derivatives (proportional to delta functions in the
coordinates normal to the potential wall) are approximately zero at this
boundary.

For this very special instance, then, I can proceed as follows

\begin{equation}
\exp -i\tau \left( K({\bf -}i{\bf \nabla })+\sum_m\eta _k\chi _{D_m}({\bf x}%
)\right) \sum_{m^{\prime }}c_{m^{\prime }}\psi _{m^{\prime }}({\bf x,}%
t)\simeq \exp (-i\tau K({\bf -}i{\bf \nabla }))\times 
\end{equation}

\begin{equation}
\exp \left( -i\tau \sum_m\eta _m\chi _{D_m}({\bf x})\right) \sum_{m^{\prime
}}c_{m^{\prime }}\psi _{m^{\prime }}({\bf x,}t)=\exp (-i\tau K({\bf -}i{\bf %
\nabla }))\times 
\end{equation}

\begin{equation}
\sum_{m,m^{\prime }}e^{-i\tau \eta _m}c_m\chi _{m^{\prime }}({\bf x})\psi _m(%
{\bf x,}t)=\exp (-i\tau K({\bf -}i{\bf \nabla }))\sum_{m,m^{\prime
}}e^{-i\tau \eta _m}c_m\delta _{mm^{\prime }}\psi _{m^{\prime }}({\bf x,}t)= 
\end{equation}

\begin{equation}
\label{14}\exp (-i\tau K({\bf -}i{\bf \nabla }))\sum_me^{-i\tau \eta
_m}c_m\psi _m({\bf x,}t)=\sum_ke^{-i\tau \eta _m}c_m\psi _m^{({\rm {free}})}(%
{\bf x,}t+\tau ) 
\end{equation}
where $\psi _m^{({\rm {free}})}({\bf x,}t+\tau )$ are wave packets
free-propagated from the original ones $\psi _m({\bf x,}t)$. (\ref{14})
plays the role of an extended impulsive approximation. It is valid only when
the wave packets I am using are such that, whenever impulsive square
potentials are activated, the $m^{{\rm {th}}}$ wave front is well within the 
$m^{{\rm {th}}}$ potential box. If I do not enforce this condition and
allow, e.g., for the waves packets to face potential walls pre-existing in
their path, then they would respond to a typical reflection model, whose
effects are completely different as we know. On the other hand, any
intermediate situation would be far more involved by using the
Campbell-Haussdorff identity.

\subsection{Supplementary conditions for the measurement of position. The
two-slit experiment}

I use now this classical experimental test in order to illustrate the
continuous need in the quantum theory of measurement for the introduction of
auxiliary hypotheses in the form of consistency conditions whenever a
particular experiment is proposed. I must expect conditions of this kind to
be associated in a inextricable way to the nature of each experiment we
design. The reason is, of course, that I am not dealing with a fundamental
theory and need to take into account whatever characteristics are imposed by
me rather than universally present. Another reason for choosing this example
is that it illustrates very well an essential feature of typical quantum
experiments, namely; the persistence of decoherence\footnote{%
Actually, the argument concerning the persistence of decoherence after a
localisation has taken place can only be accounted for once the quantum
equilibrium condition is assumed. The present one is a preliminary point to
even start talking about persistence of decoherence for, if the different
partial waves do not overlap in the course of subsequent evolution, it
doesn't even make sense to talk about decoherence, as explained here.} when
a further localisation is performed after a previous localisation has
already been carried out in the past.

In this case, the wave is a superposition of two partial waves that, after
coming through the double slit setting, at the moment of being localised in
detectors separated in such a way as to be able to discern between both
alternatives of passage can be written as:

\begin{equation}
\label{15}\psi ({\bf x},t_{{\rm {loc}}})=\psi _1({\bf x},t_{{\rm {loc}}%
})+\psi _2({\bf x},t_{{\rm {loc}}}) 
\end{equation}

But something absolutely essential for this experiment to function properly
is that both $\psi _k({\bf x,}t)$ be nonoverlapping at the time of
localisation:

\begin{equation}
\label{16}\psi _1({\bf x},t_{{\rm {loc}}})\psi _2({\bf x},t_{{\rm {loc}}%
})\simeq 0 
\end{equation}
which, somewhat more rigorously means

\begin{equation}
\label{17}\int d^3x\left| \psi _1({\bf x},t_{{\rm {loc}}})\right| ^2\left|
\psi _2({\bf x},t_{{\rm {loc}}})\right| ^2<<1 
\end{equation}

Otherwise, the experiment would not be discerning between both alternatives
of passage. The problem is that, as long as condition (\ref{16}) is
accurately satisfied, the decoherence condition

\begin{equation}
\label{18}\left| \psi ({\bf x},t_{{\rm {loc}}})\right| ^2=\left| \psi _1(%
{\bf x},t_{{\rm {loc}}})\right| ^2+\left| \psi _2({\bf x},t_{{\rm {loc}}%
})\right| ^2 
\end{equation}
is empty because, as long as both supports do not overlap, condition (\ref{18})
 and the coherent one

\begin{equation}
\label{19}\left| \psi ({\bf x},t_{{\rm {loc}}})\right| ^2=\left| \psi _1(%
{\bf x},t_{{\rm {loc}}})\right| ^2+\left| \psi _2({\bf x},t_{{\rm {loc}}%
})\right| ^2+2{\rm Re}\left( \psi _1({\bf x},t_{{loc}})\psi _2^{*}({\bf x}%
,t_{{loc}})\right) 
\end{equation}
are one and the same. Thus, it is not enough to impose a coincidence
condition to enforce decoherence effects. I need to perform a second
localisation as well, and {\it at a distance sufficiently far removed from
the first one}. This second localisation must be performed far enough from
the first one for free evolution to extend the partial waves' supports and
make both to overlap in a significant amount.

Thus, by aplying the free evolution formula so that, if $\psi ({\bf x}%
,0)=\psi _1({\bf x},0)+\psi _2({\bf x},0)$, $U(t,t_{{\rm {loc}}}+\tau )=U^{(%
{\rm {free}})}(t,\tau )$, $U(t_{{\rm {loc}}},0)=U^{({\rm {free}})}(t_{{\rm {%
loc}}},0)$, and $U(t_{{\rm {loc}}}+\tau ,t_{{\rm {loc}}})=\exp \left( -i\tau
\eta _1/\hbar \chi _{D_1}({\bf x})-i\tau \eta _2/\hbar \chi _{D_2}({\bf x}%
)\right) $ I obtain

\begin{equation}
\psi ({\bf x},t)=\left\langle {\bf x}\left| U(t,t_{{\rm {loc}}}+\tau )U(t_{%
{\rm {loc}}}+\tau ,t_{{\rm {loc}}})U(t_{{\rm {loc}}},0)\right| \psi
(0)\right\rangle = 
\end{equation}

\begin{equation}
e^{-i\tau \eta _1/\hbar }\int_{D_1}d^3x^{,}D^{({\rm {free}})}({\bf x},t;{\bf %
x}^{,},\tau )\psi _1({\bf x}^{,},0)+ 
\end{equation}

\begin{equation}
\label{19a}e^{-i\tau \eta _2/\hbar }\int_{D_2}d^3x^{,}D^{({\rm {free}})}(%
{\bf x},t;{\bf x}^{,},\tau )\psi _2({\bf x}^{,},0) 
\end{equation}

That is,

\begin{equation}
\label{20}\psi ({\bf x},t)=e^{-i\tau \eta _1/\hbar }\psi _{1,{\rm {gap}}}^{(%
{\rm {free}})}({\bf x},t)+e^{-i\tau \eta _2/\hbar }\psi _{2,{\rm {gap}}}^{(%
{\rm {free}})}({\bf x},t) 
\end{equation}
where

\begin{equation}
\label{20a}\psi _{k,{\rm {gap}}}^{({\rm {free}})}({\bf x},t)\equiv
\int_{D_k}d^3x^{\prime }D^{({\rm {free}})}({\bf x},t;{\bf x}^{\prime },\tau
)\psi _k({\bf x}^{\prime },0) 
\end{equation}
with $k=1,2$. In this approximation, both $\psi _{k,{\rm {gap}}}^{({\rm {free%
}})}({\bf x},t);$ $k=1,2$, display a time lag with respect to what their
values would be if no localising interaction had been present in their path.
It is the strong coincidence condition what restores the usual timing,
because it does not suppress the free-evolution factor from the whole
evolution operator:

\begin{equation}
\label{21}\psi ({\bf x},t)=e^{-i\tau \eta _1/\hbar }\psi _1^{({\rm {free}})}(%
{\bf x},t)+e^{-i\tau \eta _2/\hbar }\psi _2^{({\rm {free}})}({\bf x},t) 
\end{equation}

Needless to say, the interesting feature of (\ref{21}) is the fact that it
stems from approximations made on the linear and unitary quantum evolution
equation and is not a make-up.

\section{The EPR experiment for the singlet state of two identical particles.
}

When reading about nonlocality in the literature we face a nagging
difficulty not having so much to do with any real intricacy as with the
sheer ambiguity with which such term is generally burdened. An expansion of
the dictionary is not always what is needed, but here it is strictly
necessary. It seems thus feasible to look at Bell's definition of nonlocal
correlation $p(a,b)\neq p(a)p(b)$ for respective outcomes $a$ and $b$ of a
certain experiment (which are causally separated by design) as one that is 
{\it minimal} in some sense, because it rests on the more fundamental
mathematical definition of statistical dependence between two stochastic
variables. In this view, nonlocallity is implied nominally, because of what
I declare $a$ and $b$ to be. I will know this property under the quite
natural term of {\it Bell nonlocality} (BNL,) because, while it seems fair
to call it nonlocality in {\it some} sense, it is presumably weaker than
that which would be inferred from the physical picture of waves propagating
superluminically.

There is, to be sure, no actual need for the wave packets to actually
propagate faster than light to reach one another's support, for such
influence to occur. This, at least, as long as two conditions are satisfied.
Namely;

(1) The involvement of a many-particle phase space in the dynamical
equations.

(2) Entanglement between the interacting subsystems.

But it is also true that I would like to have at hand some intuitive picture
of how a mechanism so counter-intuitive in relativistic terms can take place
without any harm done to the principle of relativity. And I would like to
picture it in terms of the evolution equation of my theory. This is, of
course, no other than the Schr\"odinger equation, and work has been made on
it up to this point so that it can directly produce the desired picture.

Consider the Stern-Gerlach experiment for two neutral paramagnetic particles
in the $\frac 12\otimes \frac 12$ spin space. I will always label the
eigenstates of $\sigma _n$ with an index so I can refer without ambiguity to
a particular spin observable irrespective of the direction defining it.
Thus, e.g., $\sigma _z$ has as eigenstates $\left| +_z\right\rangle $ and $%
\left| -_z\right\rangle $, $\sigma _{{\bf n}}$ has $\left| +_{{\bf n}%
}\right\rangle $, $\left| -_{{\bf n}}\right\rangle $, etc. I also omit
unnecessary particle indexes in the spin variables, the index being implicit
in the tensor-product ordering. I will focus on the stage of the experiment
when both wave packets reach the deflection window, and I will call that
time $t_{{\rm {def}}}$ in the laboratory frame (which happens to coincide
with the CM frame). If both particles are at the (initial) moment
corresponding to the decay (the ``starting gun'' of the experiment) into a
singlet state $\left\langle {\bf x}_1,{\bf x}_2\right. \left| \Psi
(0)\right\rangle =2^{-1/2}\psi ({\bf x}_1,{\bf x}_2;0)\left( \left|
+_z,-_z\right\rangle -\left| -_z,+_z\right\rangle \right) $, after a fairly
long trip in opposite directions, they reach the configuration $2^{-1/2}\psi
({\bf x}_1,{\bf x}_2;t_{{\rm {def}}})\left( \left| +_z,-_z\right\rangle
-\left| -_z,+_z\right\rangle \right) $. At that moment both have reached
their respective windows and are simultaneously (in the LAB-frame) being
acted upon by the respective dipole term. But let us suppose the first wave
packet reaches its window slightly {\it before} the second one. We write the
corresponding interaction as $H=\eta P_{+_z}\otimes I$, then, by means of (%
\ref{coarse-grained-exponential}) with $R=\sigma _z$ and $S=\sigma _x$

\begin{equation}
\label{22}U(t_{{\rm {def}}}+\tau ,t_{{\rm {def}}})=e^{i\alpha
(z_1)}P_{+_z}\otimes P_{+_x}+e^{i\alpha (z_1)}P_{+_z}\otimes
P_{-_x}+P_{-_z}\otimes P_{+_x}+P_{-_z}\otimes P_{-_x} 
\end{equation}
with $\alpha (z)=-(\eta +z_1\Delta p_{z_1})\tau /\hbar $. The second factor
of the r.h.s. has been expanded in terms of $\sigma _x$-eigenstates in order
to see how the deflection of the first wave packet following $\sigma _z$
affects the possible values for $\sigma _x$ corresponding to the second wave
packet (we know the attempt to overcome Heisenberg's incompatibility at a
distance is the sticking point since EPR times). The action of (\ref{22}) on
the state at $t=t_{{\rm {def}}}$ is such that

\begin{equation}
\sqrt{2}\times \left| \Psi (t_{{\rm {def}}}+\tau )\right\rangle =\sqrt{2}%
\times U(t_{{\rm {def}}}+\tau ,t_{{\rm {def}}})\left| \Psi
(t_d)\right\rangle = 
\end{equation}

\begin{equation}
e^{i\alpha (z_1)}\left\langle +_x,-_z\right\rangle \left|
+_z,+_x\right\rangle +e^{i\alpha (z_1)}\left\langle -_x,-_z\right\rangle
\left| +_z,-_x\right\rangle - 
\end{equation}

\begin{equation}
\left\langle +_x,+_z\right\rangle \left| -_z,+_x\right\rangle -\left\langle
-_x,+_z\right\rangle \left| -_z,-_x\right\rangle = 
\end{equation}

\begin{equation}
\frac 1{\sqrt{2}}\left( e^{i\alpha (z_1)}\left| +_z,+_x\right\rangle
-e^{i\alpha (z_1)}\left| +_z,-_x\right\rangle +\left| -_z,+_x\right\rangle
+\left| -_z,-_x\right\rangle \right) 
\end{equation}

Now, the function I handle in QM to generate the probabilities of the
different outcomes for the observable $\sigma _z\otimes \sigma _x$ of the
composite system at time $t$ is $\left| \left\langle \left. s_z,s_x\right|
\Psi (t)\right\rangle \right| ^2$. But it is interesting to see what happens
if I calculate from this formula the marginal probabilities for the
different outcomes $s_x$ of the second particle before (making $\alpha =0$)
and after (making $\alpha \neq 0$) the first particle's $\sigma _z$%
-deflection:

\begin{equation}
\sum_{s_z}\left| \left\langle \left. s_z,\pm _x\right| \Psi (t_{{\rm {def}}%
}+\tau )\right\rangle \right| ^2=\left| \left\langle \left. +_z,\pm
_x\right| \Psi (t_{{\rm {def}}}+\tau )\right\rangle +\left\langle \left.
-_z,\pm _x\right| \Psi (t_{{\rm {def}}}+\tau )\right\rangle \right| ^2= 
\end{equation}

\begin{equation}
\label{23}=\frac 12(1\mp \cos \alpha (z_1)) 
\end{equation}

With the substitution $\alpha =0$ I recover what I would obtain if I
calculated the marginal probabilities at time $t_{{\rm {def}}}$, that is

\begin{equation}
\label{24}\left| \sum_{s_z}\left\langle \left. s_z,-_x\right| \Psi (t_{{\rm {%
def}}})\right\rangle \right| ^2=\sum_{s_z}\left| \left\langle \left.
s_z,-_x\right| \Psi (t_{{\rm {def}}})\right\rangle \right| ^2=\frac 12 
\end{equation}

The general case of $\alpha \neq 0$ I can only interpret in terms of
probabilities provided I average over the stochastic parameter $\alpha (z_1)$
(in the way previously exposed when dealing with the Stern Gerlach
experiment). Both (\ref{23}) and (\ref{24}) serve me to compare the
probabilities before ($\alpha =0$) and after ($\alpha \neq 0$) the
interaction has taken place (disregarding the effects of the free-evolution
factor from the whole evolution operator). ``Before'', I had\footnote{%
I still have to 1$^{{\rm {st}}}:$ sum the amplitudes and 2$^{{\rm {nd}}}$:
square, because I still ''do not know'' what decoherence is about. It is the
process of average, that I will introduce later more formally, what entails
the habitual decoherence rule.}

\begin{equation}
\label{25}\left| \sum_{s_z}\left\langle \left. s_z,+_x\right| \Psi (t_{{\rm {%
def}}})\right\rangle \right| ^2=0 
\end{equation}

\begin{equation}
\label{26}\left| \sum_{s_z}\left\langle \left. s_z,-_x\right| \Psi (t_{{\rm {%
def}}})\right\rangle \right| ^2=1 
\end{equation}
and ``after'', I have (provided I average to zero the trigonometric terms
appearing in (\ref{23}) and such average is represented by an overbar)

\begin{equation}
\label{27}\overline{\sum_{s_z=+,-}\left| \left\langle \left. s_z,+_x\right|
\Psi (t_{{\rm {def}}}+\tau )\right\rangle \right| ^2}=\overline{%
\sum_{s_z}\left| \left\langle \left. s_z,-_x\right| \Psi (t_{{\rm {def}}%
}+\tau )\right\rangle \right| ^2}=\frac 12 
\end{equation}
which means that the probabilities for the different outcomes of $s_x$ for
the second particle have changed as a consequence of having measured $s_z$
for the first particle even though both are causally disconnected. Now, I
new since Bell's analysis\cite{Bell} based just on quantum probabilities
(but without assuming QM dynamics) that this nonlocality was necessary
(exception made of possible loopholes). I have just presented a dynamical
discussion of how such a bizarre physical phenomenon can come about with no
need to strand into the allegedly inconsistent soil of nonlocal dynamics.
The reason can be pinned down to what I mentioned before: a combination of a
multi-particle phase space plus the involvement of entangled states. All
through this process, causality is not even touched, as the only way to
check this quantum change in the multiparticle-quantum-phase space is only
after having cropped up the results (at which time, the causality time frame
has obviously become outdated).

\section{Quantum-Stochastic-Parameter averaged \penalty-10000 Bohmian Mechanics}

I am now going to cover details that remained somewhat loose before. In
doing so, I will show why one is able to be sloppy in not considering
further entanglement between the system of interest and the rest of the
world without any harm done to the essential ideas to be extracted.
Furthermore, I will show the reason for the previous average over stochastic
phases. The idea will be presented in the context of Bohmian mechanics.

Let us consider our system ${\cal S}$, along with a second system ${\cal A}$%
, with ${\cal S}$ allowing a spectral expansion like the one giving rise to (%
\ref{coarse-grained-exponential}) and ${\cal A}$ a somewhat different one in
that we make room for an additional subspace (the one orthogonal to the rest
of the pointer states). That is, I assume the existence of a spectral
expansion for ${\cal A}$ of the form

\begin{equation}
\label{28}\sum_{q=1}^nP_q^{({\cal A})}+P^{({\cal A})\perp }=I^{({\cal A})} 
\end{equation}
while indulging the fiction that the space of states of ${\cal A}$ has
dimension $n+1$, which allows to simplify the writing, while the
corresponding generalisation would require the addition of an index\footnote{%
Such additional index would not only freight the notation, but also have a
somewhat obscure meaning, as it would not necessarility conform to any
physically sensible eigenstate expansion. The reason is the CSCO I have
chosen to expand ${\cal A}$'s space starts with the ``unnatural'' $Q^{({\cal %
A})}\equiv \sum_qq\left| {\cal A}_q\right\rangle \left\langle {\cal A}%
_q\right| $, which is fair enough to describe ${\cal A}$ in relation to $%
{\cal S}$, but not to exhaustively describe ${\cal A}$ as a system of its
own. Furthermore, the addition of a further CSCO to $Q^{({\cal A})}$, would
make me end up with a redundant description, thus risking inconsistency at
every step. The set of observables adjoined to $Q^{({\cal A})}$ in order to
reach completion is, thus, ``unnatural'' {\it by construction}.}. The
initial state of the whole system$+$apparatus adopts the form

\begin{equation}
\label{initial-state}\left| \Psi (0)\right\rangle =\sum_{q=1}^nc_q\left|
\psi _q\right\rangle \otimes \left| \Phi _0\right\rangle 
\end{equation}
where the system under experimental test is assumed to have undergone a
previous stage of preparation, and I am going to suppose the apparatus'
initial state as having the most general form possible (known as {\it %
absolute standard}):

\begin{equation}
\label{absolute-standard}\left| \Phi _0\right\rangle =\left| \Phi ^{\bot
}\right\rangle +\sum_{q=1}^na_q\left| {\cal A}_q\right\rangle 
\end{equation}

Thus, either the apparatus' transition to a pointer state as well as its
remaining on the neutral state, unaffected, are possible. We have also (for
all $q$)

\begin{equation}
\label{29}\left\langle \left. \Phi ^{\bot }\right| {\cal A}_q\right\rangle
=0 
\end{equation}

\begin{equation}
\label{30}P_q^{(A)}\left| \Phi ^{\bot }\right\rangle =0 
\end{equation}

The condition setemming from demanding normalisation for both the total
state and the $\left| {\cal A}\right\rangle $'s will be ignored. So far,
that is the general scenario. But let us be more concrete.

We recall now the standard scheme in the literature when it comes to
consider the apparatus. The following interaction Hamiltonian is proposed
for illustrative purposes

\begin{equation}
\label{31}H=Q\otimes P_y 
\end{equation}
where $P_y=-i\partial /\partial y$ is the canonical operator corresponding
to the one-dimensional pointer coordinate $y$. Then, if $\left\langle {\bf x}%
,y\left| \Psi (0)\right. \right\rangle =\sum_{q=1}^nc_{q,\lambda }\psi
_{q,\lambda }({\bf x})\Phi _0(y)$ (restoring a possible degeneracy in the
formulae,) we get

\begin{equation}
\label{32}\left\langle {\bf x},y\left| \Psi (\tau )\right. \right\rangle
=\sum_{q=1}^n\sum_\lambda c_{q,\lambda }\psi _{q,\lambda }({\bf x})\Phi
_0(y-q\tau ) 
\end{equation}

So different outputs $q$ induce different pointers in the global state. The
pointers are simply wave functions recoiled from their original position
(think of $y$ as an angle or maybe as a location in a grid). I am going to
modify slightly (\ref{31}) so that it fits the previous discussion in what
concerns the measured system's space of quantum states. That is, instead I
will consider

\begin{equation}
\label{33}H=\sum_k\eta _kP_k(Q)\otimes P_y 
\end{equation}

This amounts to changing the ``measuring rule'' $Q=\sum_qqP_q$ to a yes-no
and stochastically affected ``measuring rule'' $f(Q)=\sum_k\eta _kP_k(Q)$.
Now, in place of (\ref{32}), I would have, by applying (\ref{tensor-exponential})

\begin{equation}
\label{34}\Psi _{\left\{ \eta ,y\right\} }({\bf x},y;\tau )=\left\langle 
{\bf x},y\left| \Psi _{\left\{ \eta ,y\right\} }(\tau )\right. \right\rangle
=\sum_{k=1}^n\sum_\lambda c_{k,\lambda }\psi _{k,\lambda }({\bf x})\Phi
_0(y-\eta _k\tau ) 
\end{equation}
where $\eta $ runs over all $\eta _k$. Now, Bohmian mechanics prescribes
that in the long run, the particle guided by the quantum wave will follow
the statistical pattern of position variables ${\bf x}$ given by the modulus
of the wave function squared. But in order to be able to write a proper wave
function, I must include ${\cal A}$, so the expression to be applied the
quantum equilibrium condition to is $\left| \Psi _{\left\{ \eta ,y\right\} }(%
{\bf x},\lambda ,y;t)\right| ^2$. With this, my probability density gets
affected by the stochastics parameters $\eta $ and $y$ related to ${\cal A}$:

\begin{equation}
\rho _{\left\{ \eta ,y\right\} }({\bf x};t)\equiv \left| \Psi _{\left\{ \eta
,y\right\} }({\bf x},y;t)\right| ^2= 
\end{equation}

\begin{equation}
\label{35}\sum_{k,k^{\prime }=1}^n\sum_{\lambda ,\lambda ^{\prime
}}c_{k,\lambda }c_{k^{\prime },\lambda ^{\prime }}^{*}\psi _{k,\lambda }(%
{\bf x})\psi _{k^{\prime },\lambda ^{\prime }}^{*}({\bf x})\int dy\Phi
_0(y-\eta _k\tau )\Phi _0^{*}(y-\eta _{k^{\prime }}\tau ) 
\end{equation}

If only A had its dynamical variables fixed, I would be finished, but the
question is these variables have a stochastic character that entails some
kind of average in order to obtain the probabilities for sub-system S. In
this respect, it is very important to keep in mind that the sample space to
be averaged over is the one made up of all the $\eta $'s and all the $y$'s
(corresponding to stochastic quantum excitations that could be activated 
{\it in the same run} of the experiment). Any such average (which has to run
over both $\eta $ and $y$) will be denoted by an overbar and will be defined
in the following way

\begin{equation}
\label{36}\overline{u_{\left\{ \eta ,y\right\} }}\equiv \frac{\int d\eta
_{k_1}\cdots \int d\eta _{k_n}\int dy\pi (\eta _{k_1},\cdots ,\eta
_{k_n},y)u_{\left\{ \eta ,y\right\} }}{\int d\eta _{k_1^{\prime }}\cdots
\int d\eta _{k_n^{\prime }}\int dy\pi (\eta _1,\cdots ,\eta _n,y)} 
\end{equation}
for any function $u$ of the arguments $\eta $ and $y$, and certain
admissible density $\pi (\eta _1,\cdots ,\eta _n,y)$ for the quantum
stochastic parameters. Now, if $\Phi _0(y)$ behaves like a quasi-classical
wave function, I must expect it to oscillate very rapidly for sizable
variations of its {\it spacial} argument $y$, when the stochastic {\it energy%
} parameters $\eta _k$ also change in a characteristically macroscopic
range. The last means that the density $\pi $ must display a very slow
variation as compared to $\Phi _0$. This quasi-classical behaviour is
implemented by the vanishing of the correlation between the values of ${\cal %
A}$'s factor in the global wave function when the quantum stochastic
parameters run over their typical values and certainly depends on the
stochastic parameters' fluctuations: $\overline{\Phi _0(y-\eta _k\tau )\Phi
_0^{*}(y-\eta _{k^{\prime }}\tau )}=\delta _{kk^{\prime }}$. With this, and
provided $\pi $ is normalised, one is led to:

\begin{equation}
\overline{\left| \Psi _{\left\{ \eta ,y\right\} }({\bf x},\lambda
,y;t)\right| ^2}=\sum_{k=1}^nc_{k,\lambda }c_{k,\lambda ^{\prime }}^{*}\psi
_{k,\lambda }({\bf x})\psi _{k,\lambda ^{\prime }}^{*}({\bf x})\overline{%
\Phi _0(y-\eta _k\tau )\Phi _0^{*}(y-\eta _{k^{\prime }}\tau )}= 
\end{equation}

\begin{equation}
\label{37}\sum_{k=1}^n\sum_{\lambda ,\lambda ^{\prime }}c_{k,\lambda }\psi
_{k,\lambda }({\bf x})c_{k,\lambda ^{\prime }}^{*}\psi _{k,\lambda ^{\prime
}}^{*}({\bf x}) 
\end{equation}
which is decoherent {\it with respect to the coarse-grained} $P_k(Q)$, but
displays interferences with respect to the $\lambda $ variables to which the
experiment is blind (as obtained from the usual convention implied in the
projection postulate). Attention must be paid to the words {\it with respect
to} {\it the coarse-grained} $P_k(Q)$, because decoherent or not depends on
what observable I am considering to measure next. Thus, if I make up my mind
to use the outcoming state (\ref{37}) as an {\it incoming} state in a
further experiment designed for measuring say, $R$ with $\left[ Q,R\right]
\neq 0$ then, being $W$ diagonal in $Q$ {\it thus} not in $R$, {\it thus}
not in a new $R$-dependent coarse-graining $P_j(R)$, I find myself involved
with coherences again.

\section{Concluding remarks}

Let us briefly review the whole idea of the present work: If I want a
measurement to be ``perfect'', it must neither blur nor biass the
probability pattern of the collectivity $p({\bf x})$ previous to the
measurement. But if this is to be true and the quantum equilibrium condition
is satisfied: $p({\bf x})=\left| \sum_\lambda \left\langle {\bf x},\lambda
\left| \psi \right. \right\rangle \right| ^2$ (while retaining linear and
unitary evolution for the quantum waves,) any operator implementing this
quantum change must be diagonal in ${\bf x}$. This, in turn, makes
inescapable the form $\sum_{{\bf x}}e^{-i\eta _{{\bf x}}\tau }P_{{\bf x}}$
of which the coarse-grained version previously introduced is simply a rough
approximation suitable for use in the position representation but related to
a more general observable $Q$. Finally, Bohmian mechanics provides us with
the pointers, while a reasonable average over the stochastic parameters
involved produces the familiar coherence-loss result.

But, on the mathematical side, this is basically a work about the
exponential operator. If the handling of the exponential evolution operator
is to be trusted in any quantum mechanical context from molecular physics to
QCD, I am placing the bet in that its form and properties must also have
much to say in connection with this long-standing problem. The route
followed in this respect is little short of compelling, the only thing to be
contended being the roughness of the different approximations. More
speculative is the assumption that Bohmian mechanics be the alternative to
be followed when it comes to introducing additional variables.

But, irrespective of the final verdict of nature about Bohmian mechanics,
there are several common assertions about the theory that are in sore need
of being rebutted. One of them is the idea that Bohmian mechanics is no good
because it is ``nonlocal''. That ordinary CQM {\it is} already nonlocal
because of (\ref{collapse}) should be enough to bash away this popular
opinion forever. But there is more and is well-known long ago: nonlocality
is inferred from quantum probabilities (provided, in turn, by quantum
waves). In consequence, we should expect that it did not depend on the
particular model of additional variables we introduce. This peculiar quantum
or ``weak'' nonlocality (BNL) must be explained, and not argued against. We
should look for a picture of it in wave dynamics, and not in the
additional-variable level. That is exactly the program I have followed in
the present work.

As to the various theorems periodically launched in order to dispossess any
additional-variable model from its consistency claims, it must be said that
any such attempt must be aimed at whatever Bohmian mechanics does not
succeed to explain, not at phenomena that Bohmian mechanics {\it has no
difficulty at all} to explain. I mean, of course, spin measurements.
Actually, there is no need for a model of additional variables to determine
any pre-existing values of an arbitrary projection of spin, or of any other
``inner'' quantum variable for that matter, {\it previous} to the
measurement (``{\it beables do not need to be EPR's elements of reality!}%
''). Curiously enough, one of Bohr's famous admonishments offers a helping
hand to the additional-variable idea in this concern: it is the stochastic
nature of the measurement process which, (in conjunction with the
additional-variables' evolution, that is) leads to one or other particular
outcome for the selected quantum observable. Thus, {\it a unique} initial
value for all the dynamical variables (a particular position for the
particle and as a result, a given velocity, plus a unique initial quantum
state, plus a unique initial value for the potential energy) leads to {\it %
one or other} particular outcome {\it depending} {\it on} what particular
deflecting dipole-term we decide to introduce next and on the precise value
of the stochastic parameters involved. This is possible in the face of
K\"ochen-Specker's rigorous result, because Bohmian mechanics qualifies as
an inherently {\it contextual} additional-variable model (a possibility that
the K\"ochen-Specker theorem does not preclude). But the key to the
catch-word ``contextual'' (which otherwise would be nothing but a fancy
philosophy-laden name,) lies in the existence of discontinuous and
stochastic changes in the environment. That any quantum-orthodoxy advocate
would find unpalatable such discontinuity and stochasticity while accepting
both features in the shameful form expressed in (\ref{collapse}) would be
ironical to the extreme. However, while it is true that I have not proved
the mentioned discontinuity as an inevitable consequence of the quantum
evolution equation in the present work, I have presented fairly plausible
arguments for its ocurrence: assemblies of many-particle systems with
infinitely many quantum states as dynamical building blocks of the quantum
context giving rise to discontinuous patterns of evolution is, to say the
least, not impossible.

Another objection frequently presented to Bohmian mechanics is that its
inception in physics does not justify the effort invested when CQM already
gets the same results with much less effort. It is a question here of
deciding between mathematical simplicity and logical consistency; but
logical consistency should never be sacrificed on behalf of mathematical
simplicity (this point has been addressed already in a footnote and is
certainly not new in physics).

However, the most serious problem Bohmian mechanics has to face is that,
probably, it is not the whole story. It is much too naive to be the whole
story. This is because theories with particles are especially awkward when
it comes to discuss symmetries (this is known since the demise of Lorentz's
theory of the electron (of which Dirac's theory was a further sophistication
in a similar spirit) followed by unfruitful attempts mainly by Poincar\'e to
overcome the difficulties with a model of spacially extended electron with
``inner'' cohesion forces). Symmetries are so much better discussed in terms
of fields, and we know symmetries are an essential ingredient of modern
fundamental theories. In this respect, there is a possibility that Bohmian
mechanics be but a simpleminded model of a highly nontrivial {\it nonlinear}
field theory in which symmetries are obvious. This would-be field theory
splitting, in turn, into two separate modes of evolution: one of them made
up of stable lumps and the other made up of Schr\"odinger waves. This
possibility would presumably have to face the difficulty of sorting out
problems with renormalizability in the relativistic version and seems quirky
in that linear waves affect lumps, and not the other way about. But another
more plausible possibility is that Bohmian mechanics be a ``pointlike
version'' of a more fundamental theory of strings, branes or extended
objects in general (if not for additional location variables more concrete
than the location furnished by a dispersing wave, what sense does it make to
speak of any kind of {\it spatial} substructure beyond the level of quantum
waves?...or are strings and branes to be nothing but ``stringy'' or
``brany'' quantum numbers, ``surrealistically'' related to localisation
within the wave?). Neither of both suggestions, of course, are logical
necessities and either could be plagued by technical difficulties, but they
stress the point that Bohmian mechanics, peculiar though it is, is probably
analyzable in more cogent terms than its crude original form seems to impose.

\end{document}